\documentclass[a4paper,10pt,twoside]{cpc-hepnp}

\usepackage{multicol}
\usepackage{graphicx}
\usepackage{booktabs}
\usepackage{amssymb,bm,mathrsfs,bbm,amscd}
\usepackage[tbtags]{amsmath}
\usepackage{lastpage}

\begin{document}

\fancyhead[c]{\small Submitted to "Chinese Physics C"} \fancyfoot[C]{\small 010201-\thepage}

\footnotetext[0]{Received XX January 2014}

\title{Theoretical analysis of BLM system for HLS II\thanks{Supported by National Natural Science Foundation of China (11175180 and 11175182)}}

\author{%
      Chen Yukai(³ÂÔ£¿­)$^{1)}$\email{ykchen@mail.ustc.edu.cn}%
\quad Li Yuxiong(ÀîÔ£ÐÜ)$^{2)}$\email{lyx@ustc.edu.cn}%
\quad He Lijuan(ºÎÀö¾ê)
\quad Li Weimin(ÀîΪÃñ)
}
\maketitle

\address{%
National Synchrotron Radiation Laboratory, University of Science and Technology of China, Hefei, Anhui 230029, P.R.China\\
}

\begin{abstract}
Hefei Light Source (HLS) is being upgraded to HLS II. Its emittance will be much lower than before, therefore the Touschek scattering will increase significantly and become the dominant factor of beam loss. So it is necessary to build a new beam loss monitoring (BLM) system differed from the old one to obtain the quantity and position information of lost electrons. This information is useful in the commissioning, troubleshooting and beam lifetime studying for HLS II. This paper analyzes the distribution features of different kinds of lost electrons, introduces the new machine's operation parameters and discusses the way to choose proper monitoring positions. Base on these comprehensive analysis, a new BLM system for HLS II is proposed.
\end{abstract}



\footnotetext[0]{\hspace*{-3mm}\raisebox{0.3ex}{$\scriptstyle\copyright$}2013
Chinese Physical Society and the Institute of High Energy Physics
of the Chinese Academy of Sciences and the Institute
of Modern Physics of the Chinese Academy of Sciences and IOP Publishing Ltd}%

\begin{multicols}{2}

\section{Introduction}

For an electron storage ring, beam loss is inevitable during running period. It reduces the beam current and shortens the beam lifetime. Hence a BLM system is needed to evaluate the condition of beam loss and help researchers to optimize parameters of the ring.

DESY built the first dedicated BLM system for the electron ring of HERA in 1993\cite{lab1}. There are 214 PIN-photodiodes beam loss monitors mounted on the inner side of the vacuum chamber along the ring. This system is used to monitor the position of heavy localised beam loss and thus realize a higher current with a stable lifetime.

In HLS, we also applied a BLM system to study beam lifetime\cite{lab2,lab3}. We mounted monitors in pairs correspondingly on the inner side and the outer side of the vacuum chamber of the ring. Different signals from these two sides can help to distinguish the contribution of Touschek scattering or residual gas scattering to the beam lifetime.

HLS is now being upgraded to HLS II. Beam parameters of the new ring are significantly different from those of the old one. In order to get higher synchrotron radiation brilliance, the emittance is much lower than before. This change will increase the probability of Touschek scattering and therefore the Touschek lifetime $\tau_T$ will be the dominant factor of the beam life. In this case, we are going to build a new BLM system and enhance its ability for Touschek scattering monitoring.

\section{Three kinds of beam lifetime}

The total beam lifetime ¦Ó can be mainly divided into Touschek lifetime $\tau_q$, vacuum lifetime $\tau_v$ and quantum lifetime $\tau_T$. Their relationship can be described as  Eq.~(\ref{eq1}).

\begin{eqnarray}
\label{eq1}
\frac{1}{\tau}=\frac{1}{\tau_{q}}+\frac{1}{\tau_{v}}+\frac{1}{\tau_T}.
\end{eqnarray}

These three kinds of beam lifetime are determined by three different causes, which may lead to different distributions of lost electrons. It means that the three causes can be distinguished by the different distributions of lost electrons.

Quantum lifetime $\tau_q$: Density of the electrons in a bunch can be described by Gaussian distribution. If the electrons' momentum deviation exceeds the momentum acceptance, they will be lost. After these electrons are lost, the quantum effect will inspire the other electrons to change their momentum so that the density of the electrons can keep Gaussian distribution. Inevitably, there are always a few electrons exceed the momentum acceptance and get lost. The beam lifetime determined by this effect is named quantum lifetime. Because the Gaussian distribution is symmetrical, the lost electrons on different sides of the vacuum chamber will be symmetrical too.

Vacuum lifetime $\tau_v$: Although the vacuum pressure can reach $10^{-10}$ Torr, there are still residual gas molecules in the storage ring chamber. If the high energy electron is scattered by these residual gas molecules, its momentum may be changed. When the momentum deviation exceeds the momentum acceptance, it will be lost sooner or later. The beam lifetime determined by this effect is named vacuum lifetime. Because this effect always reduces the momentum of electrons, when these electrons pass the dipole magnet, their deflection angles will be larger than those of the electrons with nominal momentum. So the residual gas scattering will be more likely to cause electrons lost on the inner side of the ring after they pass the dipole magnet.

Touschek lifetime $\tau_T$: the transverse momentums of the electrons in the same bunch are different from each other. Since elastic scattering may occurs among these electrons, their transverse momentum may be transferred into longitudinal momentum. If one electron's longitudinal momentum increases, there must be another electron's longitudinal momentum decreasing at the same time. When the longitudinal momentum deviation exceeds the momentum acceptance, both of these two electrons will be lost\cite{lab4}. The beam lifetime determined by this effect is named Touschek lifetime. The electron which decreases longitudinal momentum will be more likely to be lost on the inner side of the ring after they pass the dipole magnet, while the other electron which increases longitudinal momentum will be more likely to be lost on the outer side of the ring.

\section{Installing way of detectors}

The BLM system is mainly used to diagnose the vacuum leakage, so the detectors are mounted on the inner side of the ring only. But the above analysis shows that the distributions of lost electrons corresponding to three different kinds of beam lifetime are different. Therefore we try to build a new BLM system which can not only diagnose the vacuum leakage but also distinguish these three beam lifetimes' contribution to the total beam lifetime based on the electrons' distributions on different sides.

In BLM system for HLS II, we plan to mount detectors in pairs on the inner side and the outer side of the vacuum chamber in horizontal direction as Fig.~\ref{fig1} and on the upper side and the lower side in vertical direction as Fig.~\ref{fig2}. This installing way of detectors can provide us a more comprehensive view of beam loss.

For example, Touschek scattering will make the electrons loss on the inner side and the outer side of the ring equivalently, while the residual gas scattering will just make the electrons loss on the inner side of the ring. So the data of lost electrons on the outer side can represent the Touschek lifetime, and the data of lost electrons on the inner side minus that on the outer side can represent the vacuum lifetime. Moreover, the detectors on the upper side and the lower side, which have not been taken into account in the HLS BLM system, will provide much more information of beam loss.

\begin{center}
\includegraphics[width=6cm]{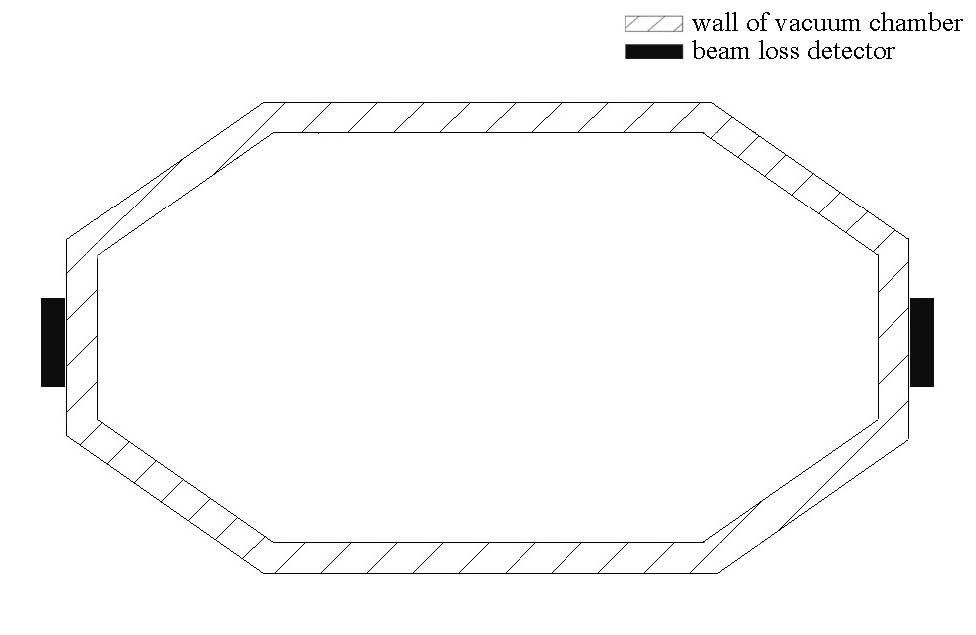}
\figcaption{\label{fig1}   Cross section of the vacuum chamber and detectors in horizontal direction. }
\end{center}

\begin{center}
\includegraphics[width=6cm]{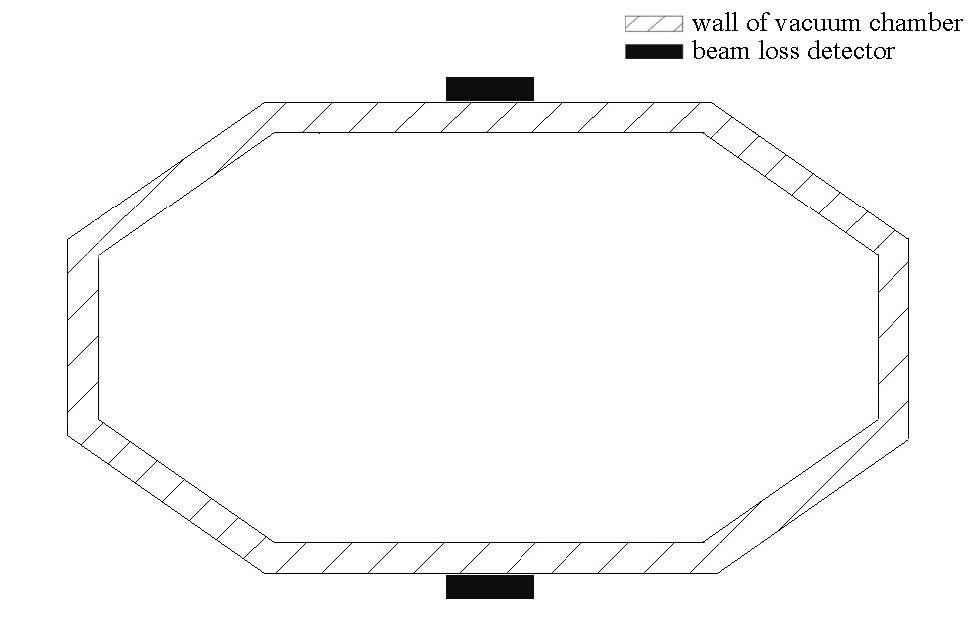}
\figcaption{\label{fig2}   Cross section of the vacuum chamber and detectors in vertical direction. }
\end{center}

\section{Monitoring positions selection}

\subsection{Principle of choosing the monitoring positions}

As the number of detectors is limited, we have to choose the most suitable locations along the ring as monitoring positions. To get high detection efficiency, we should choose where the beam loss rate is higher than other positions. And to facilitate the study of beam lifetime, we should choose where the cause of the beam loss can be distinguished by the distribution of lost electrons.

When the deviation of the electron's trajectory is greater than the limitation of the vacuum chamber, this electron will hit on the vacuum chamber wall and be lost. The greater this deviation is, the more easily the electron will be lost. So we should choose those positions, where the trajectory deviation is the greatest, as monitoring positions.

Trajectory of the electron in horizontal direction and vertical direction can be described respectively by Eq.~(\ref{eq2}) and Eq.~(\ref{eq3}).

\begin{eqnarray}
\label{eq2}
x=x_{osc}+\delta\cdot D_x=\sqrt{a_x\cdot\beta_x} cos\Phi_x+\delta\cdot D_x.
\end{eqnarray}

\begin{eqnarray}
\label{eq3}
y=y_{osc}=\sqrt{a_y\cdot\beta_y} cos\Phi_y.
\end{eqnarray}

In Eq.~(\ref{eq2}), $x_{osc}$ and $\delta\cdot D_x$ respectively represent the horizontal betatron oscillation and dispersive trajectory of the electron. The horizontal oscillation Action $a_x$ is constant for a certain electron. We can find that the larger Beta function $\beta_x$ and dispersion function $D_x$ are, the greater the trajectory deviation will be. That means we should choose those positions where the $\beta_x$ and $D_x$ are largest as the horizontal monitoring positions. Similarly, from Eq.~(\ref{eq3}), we can find that the larger $\beta_y$ is, the greater the amplitude of vertical oscillation will be. So we should choose those positions where the $\beta_y$ is greatest as the vertical monitoring positions.

\subsection{Monitoring positions of HLS II}

In order to find out the most suitable positions for HLS II BLM system by the principles mentioned above£¬it is necessary to study the lattice of HLS II. The Lattice is DBA structure in HLS II. And there are four symmetrical DBA cells in the ring of HLS II. In each cell, there are two dipole magnets, four horizontal focusing quadrupole magnets and four vertical focusing quadrupole magnets. By adjusting the magnetic field intensity of the DBA cells, HLS II can run in two modes. One is standard mode, in which the dispersion along the long straight section is zero (see Fig.~\ref{fig3}). The other is low-emittance mode, in which the dispersion along the long straight section is nonzero ( see Fig.~\ref{fig4}). We will discuss these two modes in the follow sections.

According to the Lattice of HLS II, we have calculated the Twiss parameters and the dispersion function of these two modes respectively. The results are showed respectively in Fig.~\ref{fig3} and Fig.~\ref{fig4}.
\begin{center}
\includegraphics[width=7cm]{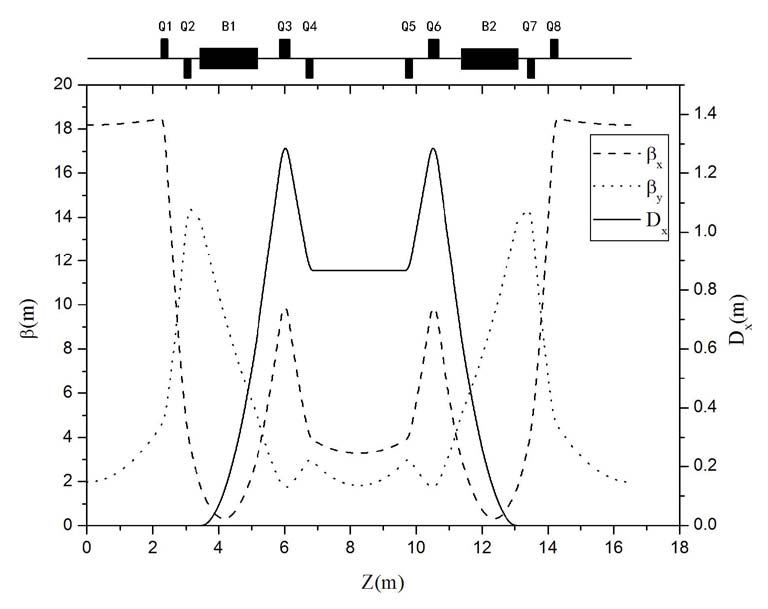}
\figcaption{\label{fig3}   Beta and dispersion function of the standard mode. }
\end{center}

Combining the distribution of the Twiss parameters and the dispersion function in these two modes, we can find that $\beta_x$ and $D_x$ may reach their maximum values before each horizontal focusing quadrupole magnets, and $\beta_y$ reach its maximum value before each vertical focusing quadrupole magnets. That means electrons will be more easily lost before each horizontal focusing quadrupole magnets in horizontal direction and each vertical focusing quadrupole magnets in vertical direction.
\begin{center}
\includegraphics[width=7cm]{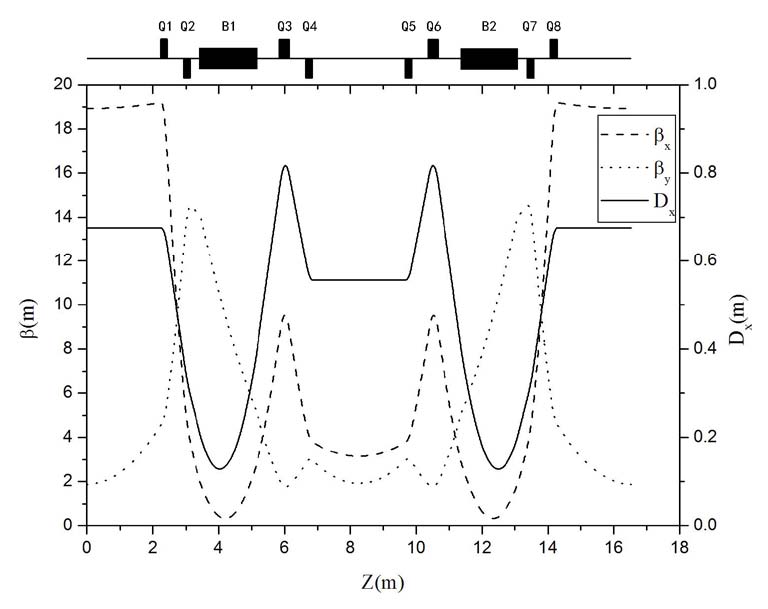}
\figcaption{\label{fig4}   Beta and dispersion function of the low-emittance mode. }
\end{center}

Based on the comprehensive analysis above, we choose those positions before horizontal focusing quadrupole magnets named Q1, Q3, Q6, Q8 as the horizontal monitoring positions and those positions before vertical focusing quadrupole magnets named Q2, Q4, Q5, Q7 as the vertical monitoring positions for HLS II BLM system. All the monitoring positions along the ring are showed in Fig.~\ref{fig5}. Detectors are mounted in pairs in horizontal direction as is showed in Fig.~\ref{fig1} at positions of H\_1 to H\_16, and in vertical direction as is showed in Fig.~\ref{fig2} at positions of V\_1 to V\_16.

\begin{center}
\includegraphics[width=7cm]{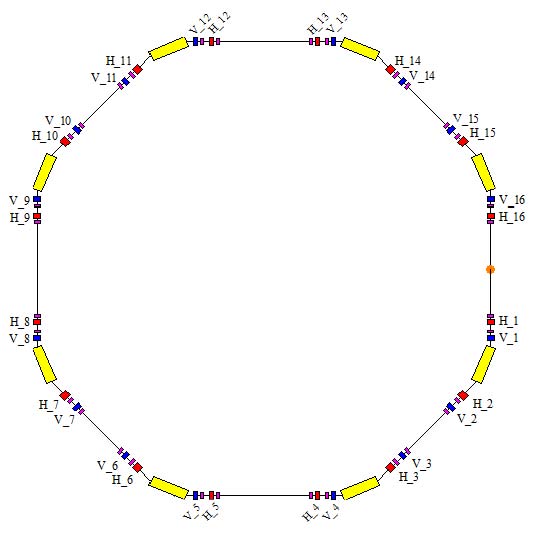}
\figcaption{\label{fig5}   (color online) Monitoring positions along the ring of HLS II. }
\end{center}

\subsection{Finial adjusting for the monitoring positions}

Since there are many other types of equipment installed along the ring, there may be no enough space for us to mount the detector at the exact position we have chosen above. Therefore we have to adjust the monitoring position according to the actual conditions. The distribution of lost electrons near the maximum point of trajectory deviation can help us to decide how to choose the new monitoring position.

\begin{center}
\includegraphics[width=8cm]{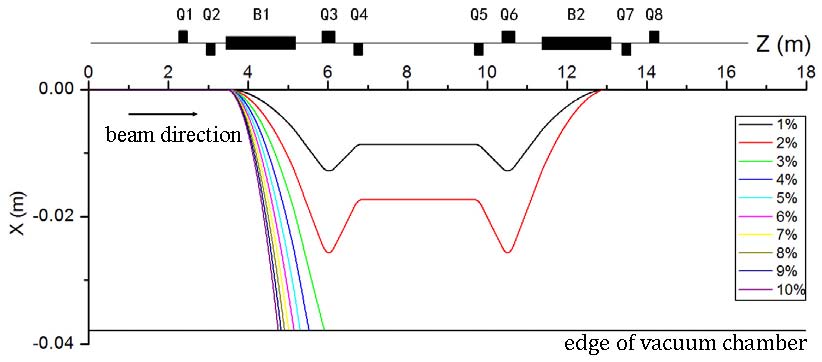}
\figcaption{\label{fig6}   (color online) Trajectory of the electron in horizontal direction with 1 to 10\% momentum deviation. }
\end{center}

For example, in the standard mode, the trajectory of the electron with 1 to 10\% momentum deviation is showed as Fig.~\ref{fig6}. We can see from the figure that the maximum momentum deviation, which is restricted by the size of the vacuum chamber, is about 3\%. If the momentum deviation is larger than 3\%, the electron will hit on the wall of vacuum chamber and get lost when it passes the dipole magnets because of excessive deflexion. The larger the momentum deviation is, the more the electron loss position will move backward to the opposite direction of the beam. And there will be few lost electrons after the maximum point of trajectory deviation. So we should adjust the monitoring position to those place before, but not after, the maximum point of trajectory deviation if it's necessary.

\section{Conclusion}

In this paper, we discuss three kinds of beam loss and the distribution of lost electrons on different sides of vacuum chamber. According to these discussing, we plan to mount detectors in pairs on the inner side and the outer side, as well as the upper side and the lower side of the vacuum chamber. This installing way of detectors can provide a more comprehensive view of beam loss, which is very useful in beam lifetime study. And monitoring positions have been chosen for HLS II, base on comprehensive analysis of Twiss parameters, dispersion function and the trajectory of electrons. The hardware and software for this new BLM system is being developed. It will play an important role in commissioning, troubleshooting and beam lifetime studying of the new ring.

\end{multicols}

\vspace{-1mm}
\centerline{\rule{80mm}{0.1pt}}
\vspace{2mm}

\begin{multicols}{2}

\end{multicols}

\clearpage

\end{document}